\documentclass[sn-mathphys,Numbered]{sn-jnl}

\graphicspath{ {} }

\usepackage{graphicx}%
\usepackage{multirow}%
\usepackage{amsmath,amssymb,amsfonts}%
\usepackage{mathrsfs}%
\usepackage[title]{appendix}%
\usepackage{xcolor}%
\usepackage{textcomp}%
\usepackage{manyfoot}%
\usepackage{booktabs}%
\usepackage{algorithm}%
\usepackage{algorithmicx}%
\usepackage{algpseudocode}%
\usepackage{listings}%
\usepackage{amssymb}
\usepackage[utf8]{inputenc}
\usepackage[T1]{fontenc}
\usepackage{diagbox}
\usepackage{array}
\usepackage{subcaption}

\usepackage{xcolor}

\begin{document}

\title[Article Title]{Crowdsourced human-based computational approach for \textcolor{black}{tagging} peripheral blood \textcolor{black}{smear sample}  images from Sickle Cell Disease patients using non-expert users}


\author[1]{José María Buades Rubio}\email{josemaria.buades@uib.es}
\author[1,2]{Gabriel Moyà-Alcover}\email{gabriel.moya@uib.es}
\author[*,1,2]{Antoni Jaume-i-Capó} \email{antoni.jaume@uib.es}
\author[1]{Nataša Petrović}\email{npe785@uib.es}

\affil[1]{UGiVIA Research Group, University of the Balearic Islands, Dpt. of Mathematics and Computer Science, 07122 Palma (Spain)}

\affil[2]{Laboratory for Artificial Intelligence Applications (LAIA@UIB), University of the Balearic Islands, Dpt. of Mathematics and Computer Science, 07122 Palma (Spain)}
\affil[*]{Corresponding author}


\abstract{In this paper, we present a human-based computation approach for the analysis of \textcolor{black}{peripheral blood smear (PBS) images} images in patients with Sickle Cell Disease (SCD). We used the Mechanical Turk microtask market to crowdsource the labeling of \textcolor{black}{PBS} images. We then use the expert-tagged erythrocytesIDB dataset to assess the accuracy and reliability of our proposal. Our results showed that when a robust consensus is achieved among the Mechanical Turk workers, \textcolor{black}{probability of error is very low, based on comparison with expert analysis}. This suggests that our proposed approach can be used to annotate datasets of \textcolor{black}{PBS} images, which can then be used to train automated methods for the diagnosis of SCD. In future work, we plan to explore the potential integration of our findings with outcomes obtained through automated methodologies. This could lead to the development of more accurate and reliable methods for the diagnosis of SCD.}


\maketitle

\section{Introduction}
\label{}
Supervised machine learning methods rely on tagged training data~\cite{orting2019survey}. The more tagged training data that is available, the more accurately the model can learn to recognize patterns and generalize to unseen data.

Crowdsourcing and Human-Based Computation (\textcolor{black}{HBC}) has become an increasingly popular approach for acquiring training labels in machine learning classification tasks, as it can be a cost-effective way to share the labeling effort among a large number of annotators. This approach can be particularly useful in cases where expert labeling is expensive or not feasible, or where a large amount of labeled data is needed to train a machine learning model~\cite{ruiz2023probabilistic}. There exist various tactics for human users to contribute their problem-solving skills~\cite{quinn2011human}:

\emph{Altruistic contribution}: This strategy involves appealing to the altruistic nature of individuals willing to contribute their time and skills to solve problems for the common good~\cite{raddick2010galaxy,kawrykow2012phylo,pharoah2014cell}.

\emph{Gamification}: This strategy involves creating engaging and fun video games incorporating problem-solving tasks~\cite{schwamb2013planet,luengo2012crowdsourcing,mavandadi2012distributed}.

\emph{Forced labor}: This strategy involves forcing website users to perform a task if they want to use its services~\cite{von2003captcha,mccoy2012development}.

\emph{Microtask markets}: This strategy involves breaking down complex tasks into smaller, simpler tasks and then outsourcing them to a large group of people~\cite{nguyen2012distributed,wang2011fusion}.

\textcolor{black}{Sickle Cell} Disease (SCD) is a serious inherited blood disorder that affects millions of people worldwide. The disease is caused by a \textcolor{black}{mutation in the HBB gene, which codes for one of the components of the hemoglobin protein}, which produces abnormal hemoglobin molecules that can cause the \textcolor{black}{Red Blood Cells (RBCs)} to have the shape of a sickle or half-moon instead of the smooth, circular shape \textcolor{black}{as normal RBCs have}~\cite{petrovic2020sickle}. \textcolor{black}{According to data from the World Health Organization (WHO) \cite{who}, it is estimated that approximately 5\% of the global population possesses the genetic traits associated with haemoglobin disorders, primarily SCD and thalassaemia. Furthermore, more than 300,000 infants born annually are afflicted with severe haemoglobin disorders. Globally, SCD resulted in 112,900 fatalities in 1990, 176,200 fatalities in 2013, and 55,3000 fatalities in 2016, as reported in previous studies~\cite{abubakar2015global,naghavi2017global}.}

\textcolor{black}{Morphological analysis of Peripheral Blood Smear (PBS) is a vital diagnostic aid for SCD.  \textcolor{black}{PBS} cannot be used for diagnosing newborns (due to sickling of cells not occurring until the baby is a bit older and switches from producing hemoglobin F to hemoglobin A), which is actually the optimal time of diagnosing SCD. It is thus only suitable for diagnosing older babies/children and adults, but also useful for monitoring treatment outcomes of already diagnosed patients. However, it is a labor-intensive and time-consuming process, which can lead to delays in diagnosis and treatment. To address this issue, automated methods for analyzing blood samples are developed, which use image analysis and machine learning algorithms to detect and count \textcolor{black}{sickle cells}~\cite{gonzalez2015red,alzubaidi2020deep,bushra2021paediatric}.
Due to this demanding and prolonged process, there is limited public availability of tagged \textcolor{black}{PBS} datasets from patients with SCD}~\cite{petrovic2020sickle,delgado2020diagnosis,gonzalez2015red,asakura1996percentage,acharya2018identification}.

We performed a systematic literature review~\cite{petrovic2020crowdsourcing} about the use of crowdsourcing \textcolor{black}{HBC} systems for the analysis of medical images. From the findings of this systematic literature review, we derived guidelines for practitioners and scientists to help them improve their research on the topic. Non-expert \textcolor{black}{HBC} for \textcolor{black}{RBC} analysis showed promising results to detect malaria parasites in digitized blood sample images~\cite{luengo2012crowdsourcing,mavandadi2012distributed} and a first attempt for SCD~\cite{jaume2016analysis}. In the literature, we also found non-expert \textcolor{black}{HBC} approaches used for labeling various types of medical images~\cite{petrovic2020crowdsourcing}, including tomographs, MRIs, retinal images, breast cancer images, endoscopic images, microscopy images, polyps, and biomarkers.
Mitry et al.~\cite{mitry2013crowdsourcing} showed encouriging results of  crowdsourcing in retinal image analysis. They achieved sensitivity of 96\% in normal versus severely abnormal detections, even without any restriction on eligible participants. Lung nodule detection with sensitivity of over 90\% for 20 patient CT datasets ~\cite{boorboor2018crowdsourcing} showed that crowdsourcing can provide highly accurate training data for computer-aided algorithms. Analysing biomedical images in ~\cite{gurari2015collect}, Gurari et al. found that after experts, non-experts performed better than algorithms and that fusing those results together yielded improved final results.

In this paper, we present an approach for the analysis of \textcolor{black}{PBS} images in patients affected by SCD through crowdsourcing \textcolor{black}{HBC} with non-expert individuals using the Mechanical Turk (MTurk) that \textcolor{black}{ is an online crowdsourcing platform that allows individuals and businesses to outsource small tasks or "Human Intelligence Tasks" to a global network of workers.} The design and experimental framework of our approach strictly adhered to the guidelines recommended by Petrovic et al.~\cite{petrovic2020crowdsourcing} in the context of crowdsourcing methodologies. Additionally, we leveraged the expert-tagged erythrocytesIDB dataset, provided by Gonzalez et al.~\cite{gonzalez2015red}, to establish the accuracy and reliability of our analysis. \textcolor{black}{We utilized the predefined categories by the dataset: circular, elongated, and other cell classifications to facilitate SCD diagnosis, as meticulously curated and labeled by medical experts, to maintain consistency with the dataset's structure, crucial for accurate analysis and cross-study comparisons. }

The aim of our research was not to substitute automated procedures utilized for diagnostic assistance in the context of patients afflicted with SCD. Instead, the main objective was to investigate the feasibility of using \textcolor{black}{HBC} to help label large datasets to facilitate the training of automated methods, particularly in situations where expert assistance is not possible. In such instances, we were chiefly interested in determining the circumstances under which we can place almost complete confidence in the labels provided by non-expert users via \textcolor{black}{HBC}.

\section{Methods and Experiments}
In this section, we propose the utilization of MTurk as a valuable tool for the analysis of \textcolor{black}{PBS} images obtained from patients with SCD. The dataset employed for this research comprised a comprehensive collection of \textcolor{black}{PBS} images derived from individuals diagnosed with SCD, obtained from a reputable medical institution. Prior to conducting the analysis, a preprocessing stage was executed to segment individual cells from full images. Subsequently, the preprocessed images were uploaded to the MTurk platform \textcolor{black}{\cite{foot1}}, where \textcolor{black}{a group of trained workers, who perform a wide range of tasks in exchange for payment, known as MTurkers, were assigned} the task of examining and annotating various properties of the \textcolor{black}{PBS} within the images. The responses collected from the MTurkers were then subjected to a quantitative measure.

\subsection{Dataset}
We used erythrocytesIDB~\cite{gonzalez2015red}, available at http://erythrocytesidb.uib.es/, which is a database of prepared blood samples from patients with \textcolor{black}{SCD}. The samples were obtained from voluntary donors by pricking their thumbs and collecting a drop of blood on a sheet. The blood was spread and fixed with a May-Grünwald methanol solution, and the images were acquired using a Leica microscope and a Kodak EasyShare V803 camera. Each image was labeled by a medical expert from "Dr. Juan Bruno Zayas" Hospital General in Santiago de Cuba, and the images were classified based on the specialist's criteria for circular, elongated, and other cells. \textcolor{black}{Examination of PBS by experienced individuals looking for features of SCD can be a sensitive test~\cite{bakulumpagi2020peripheral}}.

\subsection{Image preprocessing}
Individual cells were extracted from full images of erythrocytesIDB. The Chan-Vese active contour model~\cite{chan2001active} was employed for image segmentation. This model was chosen due to its exceptional performance in achieving a broader range of convergence and effectively handling topological changes.

The Chan-Vese method was employed without prior preprocessing steps. The application of this method resulted in the generation of a binarized image, after eliminating small objects that could potentially disrupt the subsequent classification process. We used a regularization parameter ($\mu$) value of $0.2$ and a maximum iteration limit of $1000$. However, it is noteworthy that the specified maximum iteration value was nominal, as convergence was achieved much earlier for the images under investigation.

\subsection{MTurk task design for \textcolor{black}{PBS} image analysis of patients with SCD}

The proposed approach's design and experimental framework closely followed the guidelines proposed by Petrovic et al.~\cite{petrovic2020crowdsourcing} regarding crowdsourcing methodologies. We defined a task on MTurk titled: "Sicklemia: Classify Red Blood Cells", with a description that prompts MTurkers to determine the type of \textcolor{black}{RBC}: Circular, Elongated, or Other. This task was clearly visible to MTurkers, ensuring their comprehension. It was appropriately labeled as "image, classify, red blood cells" to facilitate search and filtering based on MTurker interests.

\begin{figure}[ht!]
    \centering
    \includegraphics[width=0.8\textwidth]{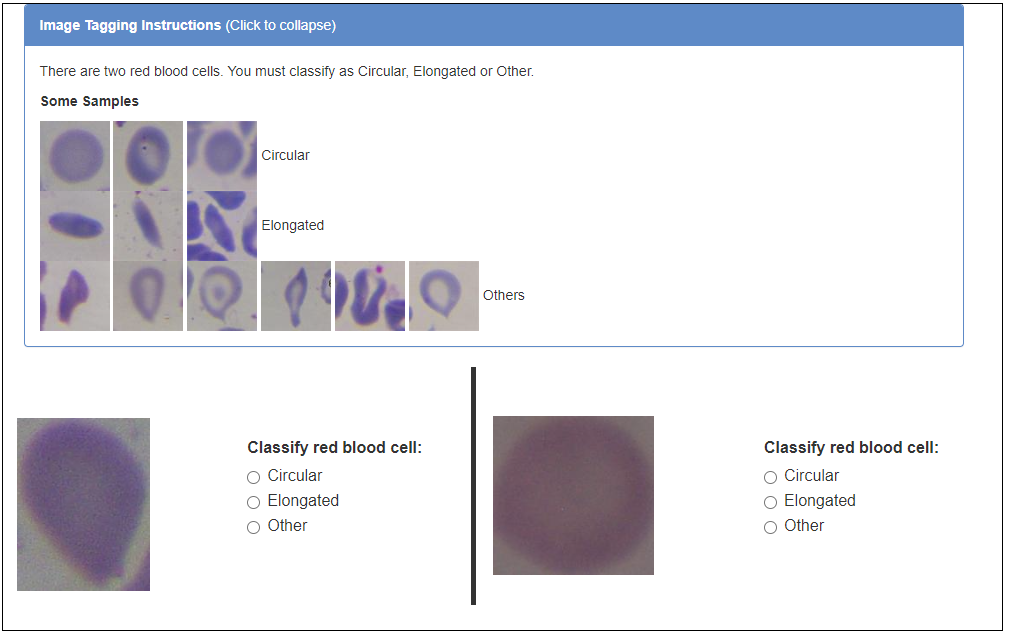}
    \caption{Instructions for cell classification.}
    \label{fig:instructions}
\end{figure}

In order to ensure a comprehensive understanding of the tasks that needed to be performed by the MTurker, a set of detailed crafted instructions was meticulously prepared. These instructions were thoughtfully designed to not only provide clear guidance but also incorporate illustrative examples for each specific task type (see Figure~\ref{fig:instructions}).

\begin{figure}[ht!]
    \centering
    \includegraphics[width=0.8\textwidth]{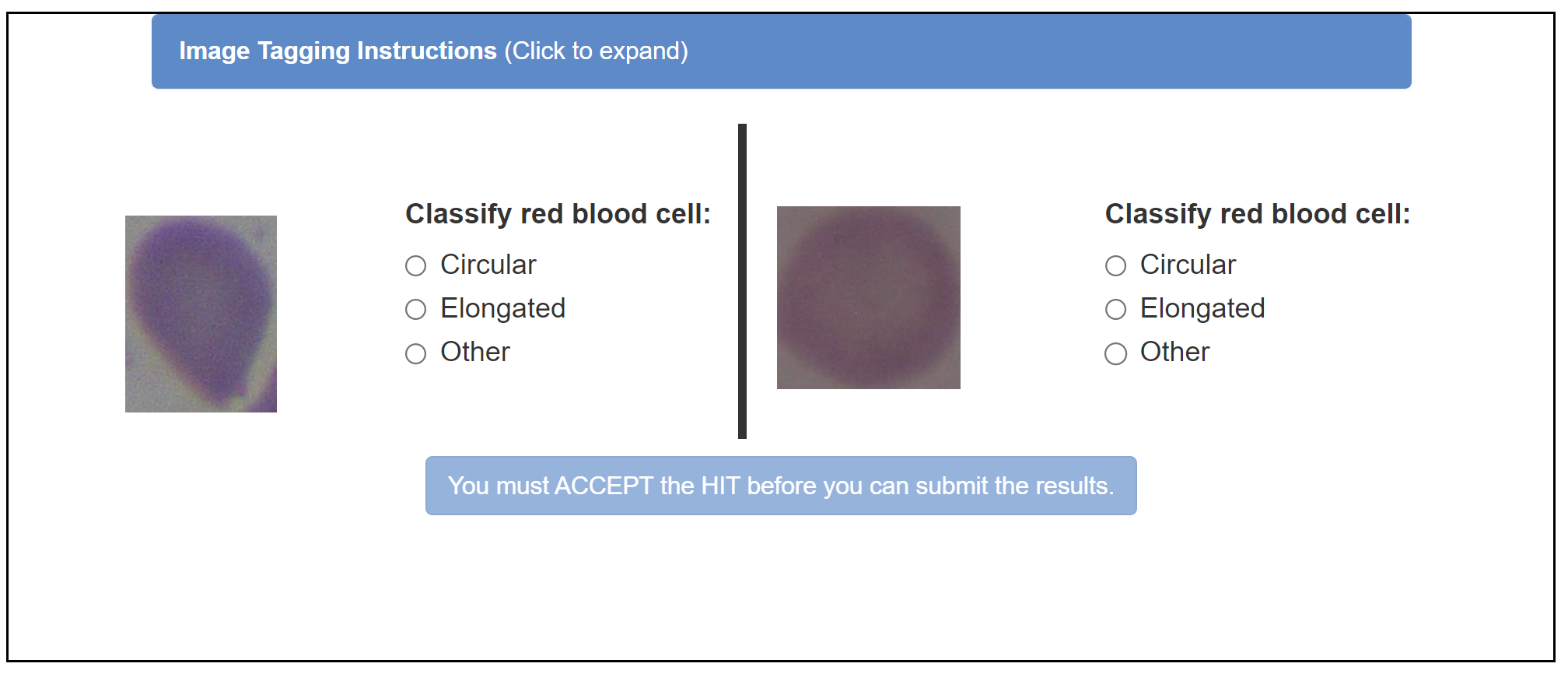}
    \caption{Cell classification task.}
    \label{fig:classify}
\end{figure}

Each MTurker was tasked with reviewing images in pairs (Figure \ref{fig:classify}). For each image pair, MTurkers were required to indicate the type of cell (Circular, Elongated, or Other). They received a reward of 0.01\$ for every classified image pair. It is important to note that not all registered MTurkers were eligible to perform these tasks, as two conditions were imposed:
\begin{itemize}
    \item Additional Requirement: Require that MTurkers be Masters to do their tasks. Master Workers on MTurk have a high success rate, holding the Masters Qualification for quality, experience, and a variety of tasks, determined through statistical analysis. 
    \item HIT Approval Rate (\%) for all Requesters' HITs greater than 90\%.
\end{itemize}

These conditions were imposed as a means of selectively filtering external MTurkers, thereby incurring a nominal cost of $0.01\$$ per processed image. Consequently, the overall cost amounts to $0.02\$$ per classified image, accounting for the multiple layers of scrutiny and assessment involved in the classification process. The requirement for each image to undergo processing by a total of five distinct MTurkers ensured a robust and reliable outcome through a collective endeavor. This multi-worker approach not only aimed to promote the \textcolor{black}{reliability} and accuracy of the classification results but also sought to mitigate potential biases or errors that may arise from relying solely on the judgment of a single worker. By harnessing the collective efforts of multiple MTurkers, the aim was to leverage diverse perspectives and expertise, thereby enhancing the overall quality and credibility of the classification process. This inclusive and collaborative approach aligns with the principles of scientific rigor and objectivity, providing a comprehensive and dependable foundation for the research findings presented in this study.

\subsubsection{MTurk parameters}
The parameters of the task were configured in order to obtain the quality of the responses needed to ensure a valid analysis and minimize the economic spending:
\begin{itemize}
\item Reward per assignment: $0.01\$$.
\item Number of assignments per task: 5.
\item Time allotted per assignment: 1 hour.
\item Task expiration period: 3 days.
\item Auto-approval and payment of MTurkers: 7 days.
\end{itemize}

\noindent MTurkers Requirements:
\begin{itemize}
\item Require MTurkers to be Masters to perform tasks: Yes.
\item Additional qualifications for MTurkers: HIT Approval Rate (\%) for all Requester's HIT greater than 90\%.
\item Task Visibility: Hidden (Only MTurkers who meet my qualification requirements can see and preview my tasks).
\end{itemize}

\subsection{Measurements}
Given a MTurker, their accuracy can be determined by comparing their responses to the \textcolor{black}{Ground Truth (GT)} \textcolor{black}{for each image, where GT is the correct and known label or category of the image.} To assess the classification performance, we \textcolor{black}{generated the confusion matrix, which is a summary of the model’s predictions versus the actual GT values, and is typically} \textcolor{black}{a square table with rows and columns representing the actual classes or categories and the predicted classes, respectively.} \textcolor{black}{We also provided raw data and calculated the Accuracy Rate and F-measure}~\cite{stapor2017evaluating, labatut2011accuracy}. We also utilized the \textcolor{black}{Sickle Cell Disease} Diagnosis Support score (SDS-score) as a measure proposed in~\cite{delgado2020diagnosis} to assess the classification of three classes \textcolor{black}{of RBCs} investigated in this study: circular, elongated cell, and other deformations. The SDS-score was designed to aid in the evaluation of SCD analysis. It was determined by calculating the ratio of the sum of true positives for all three classes to the number of sickle cells classified as other deformations and vice versa, divided by the sum of the aforementioned numerator and the sum of incorrect classifications associated with circular cells. The SDS-score indicates the usefulness of the method's results in supporting the analysis of the studied disease.

Moreover, the classification task involves imbalanced classes due to the larger quantity of circular cells compared to elongated or deformed cells. To address this issue and evaluate the overall process, we employed two measures: Class Balance Accuracy (CBA)~\cite{mosley2013,delgado2020diagnosis} and Matthews Correlation Coefficient (MCC)~\cite{Gorodkin2004,delgado2020diagnosis}. These measures provide valuable insights into the performance and effectiveness of our approach.




Regarding Accuracy Rate, for each image there were responses from $k=5$ MTurkers. If three or more responses coincided, there was a consensus and the response determined by the MTurkers was considered. The response configurations that yielded a valid response were: $5$ (complete consensus), $4-1$ (four out of the five MTurkers agreed on one class, while the remaining MTurker selected a different one, $3-1-1$ (three out of the five MTurkers agreed on one category, while each of the remaining two MTurkers selected a different one from the remaining categories), and $3-2$ (three out of the five MTurkers agreed on one class, while the remaining two MTurkers selected a different one). Otherwise, N/A (not answer) response was considered. The response configuration that did not yield a valid response was 2-2-1 (two out of the five MTurkers agreed on one class, another two MTurkers agreed on a different class, and the remaining MTurker selected yet another class). MTurkers were deemed correct if their response matched the ground truth classification.

Finally, in this study, we elucidated the methodology for computing the Accuracy Rate under the assumption of independence. Specifically, we considered the classification proficiency of a particular cell type among the MTurkers, denoting the average accuracy for this type as $\alpha$. Subsequently, we estimated accuracy ($X$) through the following procedure:

\begin{equation}    
P_X\{success\} = Acc_{estimated} = \binom{5}{1} \alpha^5 + \binom{5}{2}\alpha^4 (1-\alpha) +  \binom{5}{3}\alpha^3 (1-\alpha)^2,
\label{formula1}
\end{equation}

where first term is the case 5 \textcolor{black}{MTurkers} classify correctly, second term 4 classify correctly and the other one mistakes, and last term 3 \textcolor{black}{MTurkers} classify correctly and 2 misclassify.

\section{Results and discussion}

The accuracies for each cell type and each MTurker are detailed in Table~\ref{tab:primera}. The circular cell type demonstrated an accuracy of $86.74\%$, while the elongated and other cell types exhibited an accuracy of $67.58\%$ and $61.20\%$ respectively. Notably, when the elongated and other classes were combined into a unified category, an overall accuracy of $92.99\%$ was attained. These results highlighted the distinct accuracies associated with different cell types and underscored the enhanced performance achieved by consolidating specific categories.

\begin{table}[!htb]
\centering
\begin{tabular}{l|ccccc}
\hline
\backslashbox{\textbf{GT}}{\textbf{Prediction}} & Circular & Elongated & Other & Total & Accuracy \\ \hline
Circular                    & 2676                   & 58                     & 351                    & 3058                  & 86.74\%                       \\ \hline
Elongated                   & 48                     & 614                    & 243                    & 905                   & 67.58\%                       \\ \hline
Other                       & 69                     & 28                     & 153                    & 250                   & 61.20\%                       \\ \hline
\end{tabular}

\caption{Results of classification of each cell by each MTurker. \textcolor{black}{\textbf{GT} stands for Ground Truth}.\label{tab:primera}}
\end{table}

The adoption of a consensus-based cell type selection method, wherein a consensus was reached when 3 or more MTurkers selected the same class, produced a \textcolor{black}{improved} accuracy as shown in Table~\ref{tab:consensus}. In 20 cases there was not consensus, so the responses were considered as N/A. Notably, this approach demonstrated an overall improvement in accuracy. The results highlighted the effectiveness of leveraging consensus among multiple MTurkers to enhance the accuracy of cell type classification.

\begin{table}[!htb]
\centering
\begin{tabular}{lccc}
\hline
Consensus &  Correct & Total & Accuracy \\ \hline
Circular  & 566      & 617   & 91.73\%  \\ \hline
Elongated & 128      & 181   & 70.72\%  \\ \hline
Other     & 32       & 50    & 64.00\%  \\ \hline

\end{tabular}

\caption{Results of consensus-based cell type selection method. Consensus was reached when 3 or more MTurkers classify a cell with the same label. \label{tab:consensus}}
\end{table}

Assuming independence among the classifications, the following levels of accuracy should be obtained using the individual accuracy of 5 MTurkers, see Table \ref{tab:comparison}. The estimated outcomes exhibited superior performance compared to the observed results. This disparity challenges the assumption of independence, indicating a propensity for MTurkers to commit similar errors. These findings substantiated the inadequacy of assuming independence within the realm of MTurker behavior, underscoring the presence of correlated errors among MTurkers. The implications of these results highlighted the need for a deeper understanding of the underlying factors influencing MTurker judgments and the importance of considering inter-rater agreement in future studies.

\begin{table}[!htb]
\centering
\begin{tabular}{lcc}
\hline
Total & $Acc_{estimated}$ & Accuracy \\ \hline
Circular    & 98.11\%     & 91.73\%            \\ \hline
Elongated   & 80.73\%     & 70.72\%             \\ \hline
Other       & 70.31\%     & 64,00\%             \\ \hline
\end{tabular}

\caption{Comparison between the estimated accuracy, using Equation \ref{formula1}, and the obtained accuracy using a consensus-based cell type selection method. \label{tab:comparison}}
\end{table}

\begin{figure}[!htb]
    \centering
    \begin{subfigure}[b]{0.20\textwidth}
    \includegraphics[width=\textwidth]{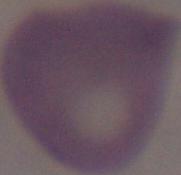}
    \caption{Other (203)}
    \end{subfigure}
    \begin{subfigure}[b]{0.20\textwidth}
    \includegraphics[width=\textwidth]{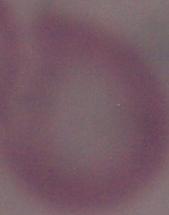}
    \caption{Other (203)}
    \end{subfigure}
    \begin{subfigure}[b]{0.20\textwidth}
    \includegraphics[width=\textwidth]{959_45__C_C2E0O3.jpg }
    \caption{NA (122)}
     \end{subfigure}
    \begin{subfigure}[b]{0.20\textwidth}
    \includegraphics[width=\textwidth]{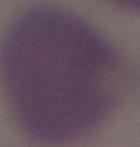}
    \caption{Other (203)}
     \end{subfigure}
    \\
    \begin{subfigure}[b]{0.20\textwidth}
    \includegraphics[width=\textwidth]{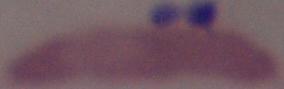}
    \caption{Other (014)}
    \end{subfigure}
    \begin{subfigure}[b]{0.20\textwidth}
    \includegraphics[width=\textwidth]{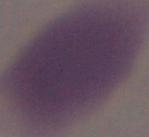}
    \caption{NA (122)}
    \end{subfigure}
    \begin{subfigure}[b]{0.20\textwidth}
    \includegraphics[width=\textwidth]{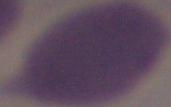}
    \caption{NA (221)}
    \end{subfigure}
    \begin{subfigure}[b]{0.20\textwidth}
    \includegraphics[width=\textwidth]{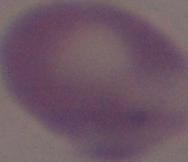}
    \caption{Circular (302)}
    \end{subfigure}
    \\
    \begin{subfigure}[b]{0.20\textwidth}
    \includegraphics[width=\textwidth]{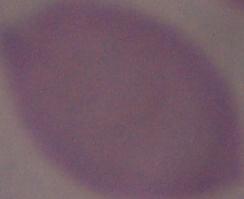}
    \caption{Circular (311)}
    \end{subfigure}
    \begin{subfigure}[b]{0.20\textwidth}
    \includegraphics[width=\textwidth]{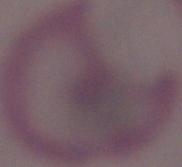}
    \caption{Circular (302)}
    \end{subfigure}
    \begin{subfigure}[b]{0.20\textwidth}
    \includegraphics[width=\textwidth]{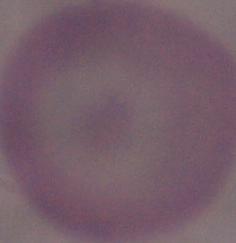}
    \caption{Circular (401)}
    \end{subfigure}
    \begin{subfigure}[b]{0.20\textwidth}
    \includegraphics[width=\textwidth]{979_44__O_C3E0O2.jpg}
    \caption{Circular (302)}
    \end{subfigure}
    \caption{MTurk miss-classifications. The top row shows circular cells, the middle row shows elongated cells, and the bottom row shows other cell types. Each label shows the class that the MTurkers have classified them, the numbers in parenthesis show the votes: circular, elongated and other. These miss-classifications are indicative of the difficulty of accurately classifying cells.}
    \label{fig:mturk_error}
\end{figure}

In Figure \ref{fig:mturk_error}, we present a collection of images showcasing instances where the MTurkers exhibit errors. The classification process employed a voting-based system, where the first row pertains to circular cell types, the second row corresponds to elongated cells, and the last row represents other cell types. The visual analysis clearly indicates the presence of challenging cases that pose difficulties for accurate classification. These observations shed light on the intricacies involved in effectively categorizing certain cell types and emphasize the importance of addressing classification uncertainties in MTurker-based studies.

Unlike computational methods, the results obtained by MTurkers provided additional information on the reliability of the decision made. This reliability was determined by the number of consensus in determining the cell's class. We separately analyzed three cases: when all 5 MTurkers agreed (463 cases), when 4 MTurkers agreed (226 cases), and when 3 MTurkers agreed (135 cases). In Table \ref{tab:metrics_3C} and Table \ref{tab:metrics_2c} we show the metrics we obtained in  these cases and compared them with the state-of-art of automated methods for analyzing blood samples  ~\cite{delgado2020diagnosis,petrovic2020sickle,asakura1996percentage,acharya2018identification, gonzalez2015red}. \textcolor{black}{Elongated and \textcolor{black}{other} cells can be consolidated because the misclassification of the normal cells as the elongated or other cells will cause the alert to the medical specialist that the patient's condition has worsened and that the therapy should be changed~\cite{delgado2020diagnosis}. Then, it is up to the specialist to review the diagnosis and to decide whether the more drastic treatment should be prescribed. This type of error is not so serious because the treatment usually has no side effects. More dangerous scenario would be to classify deformed cells (elongated or other) as normal. In this case, the specialist could decide that the patient is not at risk of a vaso-occlusive crisis, and the necessary treatment would not be applied. To support the diagnosis in a good way, classifiers need to minimize the misclassification rate of elongated cells and cells with other deformations as normal cells, and the misclassification of normal cells as elongated and cells with other deformations.} On the one hand, we can observe that if there was absolute consensus (55\% of the cases) or if 4 out of 5 MTurkers agreed (26\%  of the cases), the probability of error was very low. On the other hand, we can observe that there were only 24 cases without a consensus and 135 cases where there was consensus among 3 MTurkers, meaning these cases should be reviewed by a specialist, out of a total of 848 (19\% of the cases). 

The objective of our research is not to replace automated procedures utilized for diagnostic assistance in the context of patients afflicted with SCD. Instead, our focus is on investigating the feasibility of employing \textcolor{black}{HBC} to tag large datasets, thereby facilitating the training of automated methods, especially in situations where expert assistance is not feasible. The results demonstrate that in cases where there is a strong consensus among the MTurkers, the outcomes are comparable to the state-of-the-art automated methods. As a result, our proposed approach proves to be effective in annotating large datasets. The more tagged training data that is available, the more accurately the model can learn to recognize patterns and generalize to unseen data.



\begin{table}[!htb]
\centering
\begin{tabular}{l|rrrr}
\hline
\backslashbox{\textbf{Method}}{\textbf{Measure}} & SDS-Score & F-Measure & CBA    & MCC    \\
\hline
Delgado \emph{et al.}~\cite{delgado2020diagnosis}          & 0.95      & 0.9483    & 0.80   & 0.82   \\
Petrovic \emph{et al.}~\cite{petrovic2020sickle} GB             & 0.9518    & 0.9350    & 0.8839 & 0.8843 \\
Petrovic \emph{et al.}~\cite{petrovic2020sickle} RF             & 0.9505    & 0.9336    & 0.8806 & 0.8820 \\
Asakura \emph{et al.}~\cite{asakura1996percentage}             & 0.6180    & 0.4533    & 0.3748 & 0.3543 \\
Our proposal Individual           & 0.8759    & 0.7802    & 0.7193 & 0.6748 \\
Our proposal 5 MTurkers aggregated & 0.9009 &  0.8838  &  0.7435 & 0.7485  \\
\hline
Consensus    & 0.9272    & 0.8982    & 0.7435 & 0.7492 \\

Our proposal 5 agree          & 0.9957    & 0.9887    & 0.8571 & 0.9537 \\
Our proposal 4 agree          & 0.9204    & 0.8715    & 0.7529 & 0.7338 \\
Our proposal 3 agree          & 0.7037    & 0.6115    & 0.6108 & 0.4699 \\
\hline
\end{tabular}

\caption{Metrics obtained in the classification with 3 classes and comparison with the state-of-art. Individual, refers to results obtained from individual MTurkers. 5 MTurkers aggregated means that we consider the votes of 5 MTurkers even if the response is N/A. Consensus means that three or more MTurkers agreed on the classification.  \textit{5 agree} means that all MTurkers agreed. \textit{4 agree} means that four MTurkers agreed and one disagreed. \textit{3 agree} means that three MTurkers agreed and the other two classified differently. \label{tab:metrics_3C}}
\end{table}

\begin{table}[!htb]
\centering
\begin{tabular}{l|rrrr}
\hline
\backslashbox{\textbf{Method}}{\textbf{Measure}} & SDS-Score & F-Measure & CBA    & MCC    \\
\hline
Delgado \emph{et al.}~\cite{delgado2020diagnosis}           & 0.95      & 0.9506    & 0.89   & 0.84   \\
Petrovic \emph{et al.}~\cite{petrovic2020sickle} GB             & 0.9468    & 0.9467    & 0.9398 & 0.8872 \\
Petrovic \emph{et al.}~\cite{petrovic2020sickle} RF             & 0.9444    & 0.9442    & 0.9366 & 0.8819 \\
Acharya \emph{et al.}~\cite{acharya2018identification}             & 0.7849    & 0.7876    & 0.8116 & 0.6080 \\
González \emph{et al.}~\cite{gonzalez2015red}             & 0.4932    & 0.4897    & 0.5281 & 0.0570 \\
Our proposal Individual & 0.8759 & 0.8721  & 0.8831 & 0.7194 \\
Our proposal 5 MTurkers aggregated & 0.9009 & 0.9083  & 0.8700 & 0.7571 \\
\hline
Our proposal Consensus & 0.9272    & 0.9202    & 0.8971 & 0.8204 \\
Our proposal 5 agree     & 0.9957    & 0.9957    & 0.9868 & 0.9842 \\
Our proposal 4 agree     & 0.9204    & 0.9213    & 0.9050 & 0.8286 \\
Our proposal 3 agree     & 0.7037    & 0.7012    & 0.7131 & 0.4189 \\
\hline
\end{tabular}

\caption{Metrics obtained in the classification with 2 classes (mixing elongated and others in one class) and comparison with the state-of-art. Individual, refers to results obtained from individual MTurkers. 5 MTurkers aggregated means that we considered the votes of 5 MTurkers even if the response is N/A.  Consensus means that three or more MTurkers agreed on the classification. \textit{5 agree} means that all MTurkers agreed. \textit{4 agree} means that four MTurkers agreed and one disagreed. \textit{3 agree} means that three MTurkers agreed and the other two classified differently. \label{tab:metrics_2c}}
\end{table}

\begin{figure}[!htb]
    \centering
    \includegraphics[width=0.85\textwidth]{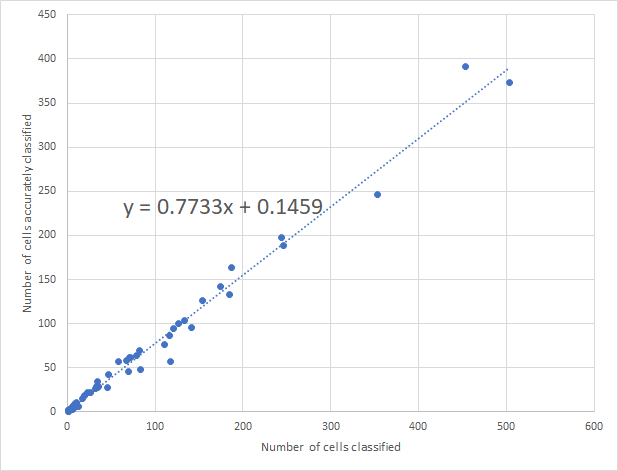}
    \caption{Ratio of cells correctly classified regarding to the number of cells classified. We can observe that this calculation can be approximated through a linear regression. The classification ratio is maintained independently of the number of classified cells.}
    \label{fig:accuracy_per_mturk}
\end{figure}

In our investigation of individual MTurkers, a notable observation emerged: an increase in the number of classifications did not yield an improvement in accuracy. This finding is visually represented in Figure \ref{fig:accuracy_per_mturk}. The results challenge the prevailing assumption that increased participation levels invariably lead to enhanced performance. These findings prompt a reevaluation of the role of quantity versus quality in the context of MTurker contributions, raising important considerations for optimizing crowd-based classification tasks.

\section{Conclusions}
This research paper introduced an approach for the analysis of Red Blood Cell images in patients afflicted by Sickle Cell Disease. The proposed method leverages crowdsourcing Human-based Computation by engaging non-expert individuals through the Mechanical Turk microtask market, especially in situations where expert assistance is not feasible.

The findings of this study indicate that when a robust consensus is achieved among the Mechanical Turk micro-task market workers, the results exhibit that \textcolor{black}{probability of error is very low, based on comparison with expert analysis}. Consequently, our proposed approach could be employed for dataset annotation.

The present study incorporates the confusion matrices, along with the raw data, within the results to facilitate researchers in computing additional metrics. The dataset utilized in this research can be accessed at http://erythrocytesidb.uib.es/. In the interest of advancing scientific knowledge, it is advantageous for authors to share their raw data and image datasets used in their investigations.

\textcolor{black}{The morphological analysis of PBS as a diagnostic tool for SCD are still used by some health systems and hospitals, even so we acknowledge recent developments in SCD point-of care diagnostics~\cite{ilyas2020smartphone}. In this work we verified that non-expert users had good results in labeling tasks for circular, elongated and other, with the aim that as further work we can tag large PBS datasets from patients with SCD with non-expert users to feed automated methods. Moreover, we consider that our method could be transferable to new cells morphologies~\cite{lynch1990peripheral} for other hemoglobinopathies that can be detected/analyzed/diagnosed by visual inspection methods. For this reason, as a further work we are interested in validating our proposal with other hemoglobinopathies.}

\textcolor{black}{Moreover, }this research endeavors to establish the fundamental principles for the effective labeling of extensive datasets, particularly in scenarios where expert involvement is unfeasible. As part of future work, it foresees explorations aimed at investigating the potential integration of these findings with outcomes obtained through automated methodologies.  Within the context of extensive dataset labeling, the incorporation of human-decided, reliable labels in conjunction with those obtained through automated methods holds notable significance. This dual-input approach has the potential to mitigate the risk of preserving errors and biases inherent in automated methods during the final labeling process. Consequently, this methodology could lead to a reduction in the transfer of such biases during the training of subsequent models, ultimately enhancing the quality of derived insights and predictive outcomes.

\section*{Ethical Approval}
not applicable

\section*{Funding}
Project PID2019-104829RA-I00 “EXPLainable Artificial INtelligence systems for health and well-beING (EXPLAINING)” funded by \\ MCIN/AEI/10.13039/501100011033. 

\section*{Availability of data and materials }
 erythrocytesIDB http://erythrocytesidb.uib.es/
 
\section*{Author Contributions}
J. B., G. M., and A.J. wrote the main manuscript text. A.J. and N.P. wrote the state of the art. J.B. prepared the data and executed the experiments. All authors designed the experimentation and reviewed the manuscript.






\bibliography{bibliography}

\end{document}